\documentstyle[stwol,epsf]{article}

\def\Journal#1#2#3#4{{#1} {\bf #2}, #3 (#4)}

\def\PLB{{\em Phys. Lett.}  B}
\def\PRL{\em Phys. Rev. Lett.}
\def\PRD{{\em Phys. Rev.} D}

\def\JPG{{\em J. Phys.} G}

\def\be{\begin{equation}}
\def\ee{\end{equation}}
\def\ba{\begin{eqnarray}}
\def\ea{\end{eqnarray}}

\begin{document}

\begin{titlepage}

\begin{center}

\renewcommand{\thefootnote}{\fnsymbol{footnote}}

\vspace{3 cm}

{\huge\bf A Separate SU(2) for the Third Family: Topflavor~\footnote{Talk 
given by S.~Nandi at the 28$^{th}$ International Conference in High Energy
Physics (ICHEP96), Warsaw, Poland, July 1996.}}

\vspace{0.7 cm}

{\Large\bf D.~J.~Muller and S.~Nandi}~\footnote{On sabbatical leave at the
University of Texas at Austin from Oklahoma State University.}

\vspace{0.7 cm}

{\Large\bf Department of Physics, Oklahoma State University, Stillwater,
OK 74078, USA \\
and \\
Department of Physics, University of Texas, Austin, TX 78712, USA}

\vspace{2,3 cm}

{\bf Abstract}

\end{center}

We consider the extended electroweak gauge group 
SU(2)$_1\times$SU(2)$_2\times$U(1)$_Y$ where the first and second families
of fermions couple to SU(2)$_1$ while the third family couples to SU(2)$_2$.
Bounds based an precision electroweak observables and heavy gauge boson 
searches are placed on the new parameters of the theory.
The extra gauge bosons can be as light as about a TeV and can 
be discovered at future colliders such as the NLC and LHC for a wide
range of the parameter space.
FCNC interactions are also considered.

\vfill

\noindent OSU Research Note 316 \\
UTEXAS-HEP-96-17 \\
DOE-ER40757-086

\end{titlepage}

\title{A SEPARATE SU(2) FOR THE THIRD FAMILY: TOPFLAVOR}

\author{D. J. MULLER, S. NANDI\,$^a$}

\address{Department of Physics, Oklahoma State University, Stillwater, 
OK 74078, USA \\
and \\
Department of Physics, University of Texas at Austin, TX 78712, 
USA\,$^b$}

\twocolumn[\maketitle\abstracts{
We consider the extended electroweak gauge group 
SU(2)$_1\times$SU(2)$_2\times$U(1)$_Y$ where the first and second families
of fermions couple to SU(2)$_1$ while the third family couples to SU(2)$_2$.
Bounds based an precision electroweak observables and heavy gauge boson 
searches are placed on the new parameters of the theory.
The extra gauge bosons can be as light as about a TeV and can 
be discovered at future colliders such as the NLC and LHC for a wide
range of the parameter space.
FCNC interactions are also considered.
}]

\section{Introduction}

We consider the possibility that the third generation of fermions have 
different electroweak interactions from the first and second. In particular, 
we consider the extended electroweak gauge group 
SU(2)$_1\times$SU(2)$_2\times$U(1)$_Y$ where the first and second generations 
of fermions couple to SU(2)$_1$ while the third generation of fermions 
couples to SU(2)$_2$. We call this model Topflavor in analogy to the Topcolor
model.\cite{Hill} Similar models of generational nonuniversality have been
considered in 
the past;\,\cite{previous} here we investigate the phenomenology of the
model. In particular, we determine the prospects for observing these extra
gauge bosons at future colliders.

\section{The Model}

We first present a brief overview of the model. The left-handed first and 
second generation fermions transform as doublets under SU(2)$_1$ and are
singlets under SU(2)$_2$. Conversely, the left-handed third generation 
fermions form doublets under SU(2)$_2$ and singlets under SU(2)$_1$. All 
right-handed fields are singlets under both SU(2)'s. With these 
representations the theory is anomaly free. The covariant derivative is
\be
D_\nu = \partial_\nu - i g_1 \stackrel {\rightarrow}{T} \cdot 
\stackrel {\rightarrow}{W_\nu} - i g_2 \stackrel {\rightarrow}{T'} \cdot
\stackrel {\rightarrow}{W'_\nu} - i \frac{g'}{2} Y B_\nu
\ee
where the $W^a$ belong to SU(2)$_1$ and the $W'^a$ belong to SU(2)$_2$.

The symmetry breaking in the model is accomplished in two stages. In the 
first stage the SU(2)$_1\times$SU(2)$_2$ is broken down at a scale 
$u \sim$ TeV to the SU(2)$_W$ of the standard model (SM). This is accomplished 
by introducing a Higgs field $\Phi$ that transforms as (2, 2, 0) under
(SU(2)$_1$, SU(2)$_2$, U(1)$_Y$).

In the second stage the remaining symmetry, SU(2)$_W\times$U(1)$_Y$, is
broken down to U(1)$_{em}$. This is accomplished by introducing 
two Higgs fields that we call H$_1$ and H$_2$. H$_1$ transforms as
(2, 1, 1) and obtains a vacuum expectation value (vev) 
$\langle {\rm H}_1 \rangle = (0, v_1)$. H$_2$ transforms as (1, 2, 1) and
develops a vev $\langle {\rm H}_2 \rangle = (0, v_2)$.\cite{topflavor}

The gauge bosons of the theory obtain mass through their interactions
 with the
Higgs fields. In the neutral sector, the fields in the current basis
($W_3$, $W'_3$, $B$) are related to the fields in the mass basis ($\gamma$,
$Z_l$, $Z_h$) by an orthogonal matrix, {\bf R}:
\be
\pmatrix{W_3 \cr W'_3 \cr B \cr} = {\rm \bf R}^\dagger (\phi, \theta_W, 
\epsilon_1, \epsilon_2) \pmatrix{\gamma \cr Z_l \cr Z_h \cr}
\ee
where $Z_l$ is called the ``light $Z$ boson'' (associated with the $Z$ boson
observed at present day colliders) and $Z_h$ is 
called the ``heavy Z boson''. Moreover, $\epsilon_1 \equiv v_1^2/u^2$, 
$\epsilon_2 \equiv v_2^2/u^2$, $\theta_W$ is the weak mixing angle, and
$\phi$ is 
an additional mixing angle such that the couplings of the theory are related
to the electric charge by $g_1 = \frac{e}{\cos \phi \sin \theta_W}$,
$g_2 = \frac{e}{\sin \phi \sin \theta_W}$, and 
$g' = \frac{e}{\cos \theta_W}$.

In the charge sector, the mass eigenstates are denoted by $W_l$ and $W_h$ 
where $W_l$ is the $W$ boson observed at present colliders and $W_h$ is
termed the ``heavy W boson''. These are related to the current basis 
($W$, $W'$) by an orthogonal matrix, ${\rm \bf R}'$:
\be
\pmatrix{W \cr W' \cr} = {\rm \bf R}'^\dagger (\phi, \epsilon_1, \epsilon_2)
\pmatrix{W_l \cr W_h \cr} \ .
\ee

Due to the enlarged gauge and Higgs structures of this model, the couplings
of the particles are modified from their SM values. These modifications are a
function of $\phi$, $\epsilon_1$, and $\epsilon_2$, but in the limit of 
$\epsilon_1 = 0$ and $\epsilon_2 = 0$, the SM couplings of the fermions are 
recovered. Due to the modification of the particle couplings and the presence 
of extra
gauge bosons, the phenomenology of this theory is different from the SM.
We investigate the case where the theory is perturbative. This 
places a 
restriction on the values that $\phi$ can take: the requirements that 
$g_2^2/4\pi < 1$ and $g_1^2/4\pi < 1$ give $\tan \phi > 0.2$ and 
$\tan \phi < 5.5$, respectively.

\section {Constraints from Experimental Data}

We restrict the possible values of the three new parameters ($\tan \phi$, 
$\epsilon_1$, and $\epsilon_2$) of the theory by requiring that the 
theoretical predictions for various processes agree with the experimental
values. 
We performed a fit to the following precision $Z$ pole observables:
$\Gamma_Z$, $R_l$, $R_\tau$, $R_c$, $R_b$, $\sigma_{had}^\circ$, 
$A_{FB}^{\circ, b}$, $A_{FB}^{\circ, l}$, $A_{FB}^{\circ, \tau}$, and
$A_{LR}^{\circ, \tau}$. The current values for $R_b$ and $R_c$ were used:
$R_b = 0.2178 \pm 0.0011$ and $R_c = 0.1715 \pm 0.0056$.\cite{Zobs}
Also included in the fit are the width and mass 
of the $W$ boson and the $\tau$ lifetime.\cite{Wtau}
To simplify the analysis, we considered various cases where the ratio
of the epsilons, $\frac{\epsilon_1}{\epsilon_2}$, is constrained to
some fixed number.

\begin{figure}[tb]
\setlength{\epsfysize}{53mm}\epsfbox[24 367 589 768]{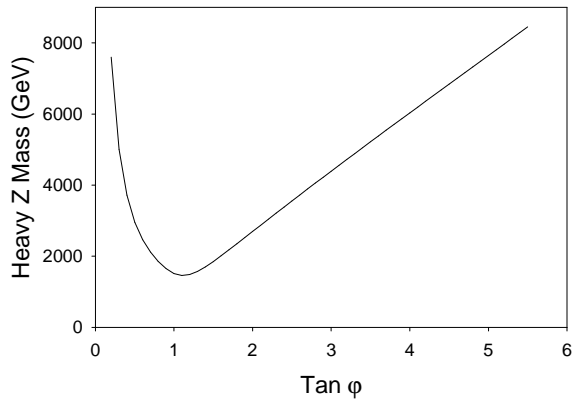}
\caption{Restriction Curve for $\frac{\epsilon_1}{\epsilon_2} = 1$.
The region above the curve gives allowed values of the parameters, while
the region below the curve is phenomenologically excluded.}
\label{fig1}
\end{figure}

\begin{figure}[tb]
\setlength{\epsfysize}{52mm}\epsfbox[14 365 591 765]{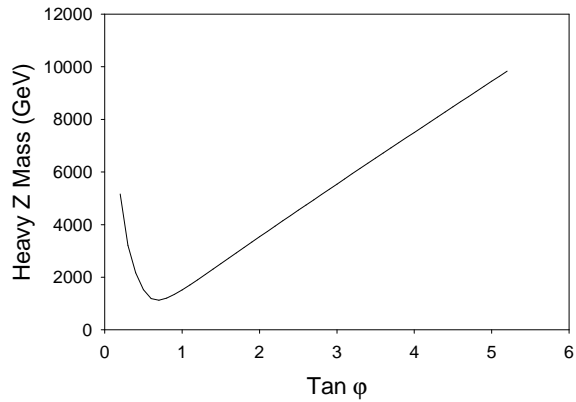}
\caption{Restriction Curve for $\frac{\epsilon_1}{\epsilon_2} = 3$.}
\label{fig2}
\end{figure}

Typical results are shown in figures~\ref{fig1} and~\ref{fig2}. In 
Fig.~\ref{fig1} we have set the constraint that 
$\frac{\epsilon_1}{\epsilon_2} = 1$, while in Fig.~\ref{fig2} 
$\frac{\epsilon_1}{\epsilon_2} = 3$.
The curves are taken at the 0.05 significance level where the region above
the curves contains
the parameters allowed by experiment while the region
below gives values for the observables inconsistent with experiment.
The smallest mass allowed in Fig.~\ref{fig1}
is 1450 GeV with $\tan \phi$ around 1.1,
while in Fig.~\ref{fig2} the lowest mass is 1120 GeV with 
$\tan \phi$ around 0.7.

In general the experimental data restrict the masses of the heavy gauge
bosons to masses greater than about 1.1 TeV. The greatest restriction 
comes from the $\tau$ lifetime.

In this model, the value of $R_b$ can be greater or less than the SM value,
so there is the possibility that this model can explain the $1.8 \sigma$ 
deviation in the experimental value of $R_b$ from the SM value. We have that
\be
\frac{\delta R_b}{R_b} = 0.8456 \cos^2 \phi 
\frac{\epsilon_1 \sin^2 \phi - \epsilon_2 \cos^2 \phi}{(g_V^b)^2 + (g_A^b)^2}
\ee
where $\delta R_b$ is the Topflavor model value of $R_b$ minus the SM value.
Thus the direction $R_b$ goes in this model depends on the interplay of
$\epsilon_1$ and $\epsilon_2$ tempered by $\phi$. The recent value of $R_b$,
$0.2178 \pm 0.0011$, can be accommodated to within $1 \sigma$. For example,
with $\frac{\epsilon_1}{\epsilon_2} = 3$, $\tan \phi = 0.9$, and 
$M_{Z_h} = 1400$ GeV, $R_b$ is 0.2169. This model provides no explanation
for the $A_{FB}^{\circ, \tau}$ anomaly.\cite{Zobs}

\section{Processes at Future Colliders}

\begin{figure}[tb]
\setlength{\epsfysize}{52mm}\epsfbox[26 385 560 774]{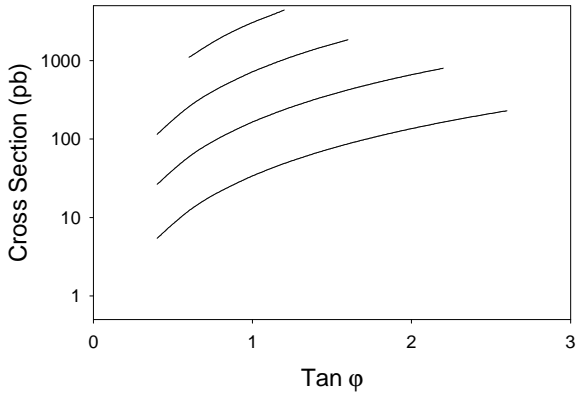}
\caption{Heavy $W$ Production at the LHC. From top
to bottom, the curves are for $M_{W_h}$ = 2000, 3000, 4000, and 5000 GeV.}
\label{fig3}
\end{figure}

We now consider the potential of future colliders to detect phenomenology
particular to this model. The best chance of observing the heavy $W$ 
boson is at a high energy hadronic collider such as the LHC. The 
optimum condition for producing $W_h$ in such a fashion is when 
$\tan \phi$ is large since the coupling of the first generation fermions 
to the $W_h$ goes as $\tan \phi$ and hence the production rate goes as 
$\tan^2 \phi$.
The cross section for the production of the $W_h$ at the LHC with 
$\sqrt{s} = 14$ TeV is shown in Fig.~\ref{fig3}. The curve takes
$\frac{\epsilon_1}{\epsilon_2} = 1$, but the results are very similar
for other values of that ratio. As the figure shows, the value of the cross
section can be quite large for smaller values of allowed mass: for example
with $\tan \phi = 1$ and $M_{W_h} = 2$ TeV, the cross section is about
3 nb. 
Thus, if the heavy $W$ boson has a mass in
the few TeV range, it will be within the discovery reach of the LHC.

At the LHC there is also the potential for single top production through
the 
production of a heavy $W$ boson and its subsequent decay to a top and bottom
quark. For allowed values of the parameters and a heavy $W$ mass in the 
few TeV range, the cross section for this process typically ranges
between a tenth to a half of what it is for top pair production through 
gluon fusion. For example, for $\tan \phi = 1$ and $M_{W_h} = 2000$ GeV, 
the cross section is 0.44 nb. A distinctive signature of this process
is 2 b-jets plus a single large $p_{\rm T}$ charged lepton.

A process of considerable current interest is $e^+ e^- \rightarrow
W_l^+ W_l^-$ which will provide an important test of the SM prediction for
the tri-gauge coupling of the $W$ to the $Z$. 
Unfortunately, the Topflavor prediction for the $W_l$ pair production cross 
section at LEP-II is not significantly different from the SM value for
allowed values of the $Z_h$ mass. This is largely due to the fact that the
coupling of $W_l$ to $Z_h$ vanishes to zeroth order in the $\epsilon_1$
and $\epsilon_2$. 

In the Topflavor model, the process $e^+ e^- \rightarrow \mu^+ \mu^-$
gets extra contributions from $Z_h$ exchange. Since the $Z_h$ mass has to
be greater than a TeV, the deviation in the predicted cross section from 
the SM value
is insignificant for LEP-II energies. At the NLC, however, the difference
in the predictions for the cross section is significant.

\section{FCNC effects}

This theory contains FCNC
interactions at tree level 
since the couplings of the third family's left-handed fermions to the $Z$'s
differ from those of the first and second families.
The part of the Lagrangian that contains the
FCNC interactions is
$$
\nonumber {\cal L}^{off-diag}_{neutral} = \overline{U}_L \gamma^\mu
X_L^\dagger
[ d_l^{(u)} Z_{l,\mu}  +  d_h^{(u)} Z_{h,\mu} ] X_L U_L
$$
\be
+ \ \overline{D}_L \gamma^\mu Y_L^\dagger [ d_l^{(d)} Z_{l,\mu}
+ d_h^{(d)} Z_{h,\mu} ] Y_L D_L
\ee \label{fcnc}
where $U_L \equiv (u, c, t)_L$ and $D_L \equiv (d, s, b)_L$. Only the (3,3)
elements of the four matrices $d_l^{(u)}$, $d_h^{(u)}$, $d_l^{(d)}$, and
$d_h^{(d)}$ are nonzero and these are given in terms of $\epsilon_1$, 
$\epsilon_2$, and $\phi$. Moreover, $X_L$ and $Y_L$ relate the gauge to 
the mass eigenstates of the up and down sectors, respectively. Since $X_L$
and $Y_L$ are related to the CKM matrix, $K$, by $K = X_L^\dagger Y_L$, we
can use the CKM matrix to eliminate $X_L$ from Eq.~\ref{fcnc}. Then using the 
observed hierarchy of the CKM matrix, $K_{ii} \simeq 1 >> K_{ij}$ for
$i \not = j$, we can relate the FCNC couplings in the up and down sectors.
For example, the part of the Lagrangian involving $Z_l$ with the top and
charm quarks can be written as
\be
\{ \overline{c}_L \gamma^\mu [ (d^{(u)}_l)_{33}
(Y_L^\dagger)_{23} (Y_L)_{33} ] t_L 
\nonumber + h.c. \} Z_{l,\mu} \ , 
\ee
while that involving $Z_l$ with the strange and bottom quarks is
\be
\{ \overline{s}_L \gamma^\mu [ (d^{(d)}_l)_{33}
(Y_L^\dagger)_{23}
(Y_L)_{33} ] b_L + h.c. \} Z_{l,\mu} \ .
\ee
We get similar expressions involving $Z_h$. These contain the same factor
of $(Y_L^\dagger)_{23} (Y_L)_{33}$, thus
the FCNC interactions 
involving $b \overline{s}$ will constrain those involving $t \overline{c}$.

As an example, we can use the limit on the the branching fraction 
$B(B_s^\circ \rightarrow \mu^+ \mu^-) < 8.4 \times 10^{-6}$ set by the CDF
collaboration \cite{CDF} to obtain an upper limit on single top production
at an NLC
type collider. For example, taking $\tan \phi = 1.0$, $M_{Z_h} = 2.2$ TeV, 
$\sqrt{s} = 1$ TeV, and $B(B_s^\circ \rightarrow \mu^+ \mu^-) \sim 10^{-6}$
gives $\sigma(e^+ e^- \rightarrow t \overline{c}) = 0.2$\,pb. One 
distinctive signature of this process is a single b-jet and a large
$p_{\rm T}$ lepton coming from the decay of the top.

\section{Conclusion}

The model presented here has extra, essentially degenerate, charged and
neutral gauge bosons which 
give rise to interesting phenomenology.
Analysis of the existing data allow the extra
gauge bosons to have mass as low as about a TeV and as low as a few TeV
 for a 
wide range of the mixing angle $\phi$. The experimental value of $R_b$ can be 
accommodated. Future colliders such as the LHC will be able to observe these gauge
bosons directly or indirectly if the mass is in the few TeV range. The model gives
rise to single top production at both hadronic and leptonic colliders with
clear signatures. There is the potential for FCNC interactions with
observable rates. 

Since one of the weak SU(2)'s can be stronger than SU(2)$_W$, this model can
have baryon number violation via an instanton channel ala
't Hooft.\cite{instanton} This will further restrict the allowed parameter space
in $\phi$. These and other details of the model will be presented 
elsewhere.\cite{future}

\section*{Acknowledgements}
We wish to thank Duane Dicus of the University of Texas at Austin for his 
hospitality and support during this sabbatical leave. This research was 
supported by a grant from the U.S. Department of Energy, Grant No.
DE-FG02-94ER40852.

\vskip 18 pt

\noindent a: Talk presented by S.~Nandi.\\
b: Present address, on sabbatical leave at the University of Texas 
from Oklahoma State University.


\section*{References}
\begin{thebibliography}{99}

\bibitem{Hill}C.~T.~Hill, \Journal{\PLB}{266}{419}{1991};
{\it ibid.}, B {\bf 345}, 483 (1995); C.~T.~Hill and S.~J.~Parke,
\Journal{\PRD}{49}{4454}{1994}.

\bibitem{previous}X.~Li and E.~Ma, \Journal{\PRL}{47}{1788}{1981};
{\it ibid.}, {\bf 60}, 495 (1988); \Journal{\PRD}{46}{R1905}{1992};
\Journal{\JPG}{19}{1265}{1993}; R.~S.~Chivukula, E.~H.~Simmons, and
J.~Terning, \Journal{\PLB}{331}{383}{1994}; {\it ibid.}, 
\Journal{\PRD}{53}{5258}{1996}; E.~Malkawi, T.~Tait, and C.~P.~Yuan,
hep-ph/9603349.

\bibitem{topflavor}A preliminary investigation of the Topflavor model with
only one Higgs doublet can be found in D.~J.~Muller and S.~Nandi, 
hep-ph/9602390, to be published in {\em Phys. Lett.} B; see also
E.~Malkawi, T.~Tait, and C.~P.~Yuan, hep-ph/9603349.

\bibitem{Zobs}See A.~Blondel, talk at
this conference;
Combination of ALEPH, DELPHI, L3, SLD, and OPAL results, quoted by Paul
Langacker and Marcel Demarteau at the 1996 meeting of the Division of
Particles and Fields, Minneapolis, MN.

\bibitem{Wtau} Particle Data Group, R.~M.~Barnett {\it et al.}, 
{\em Phys. Rev.} D54 (1996).

\bibitem{CDF}CDF Colaboration, F.~Abe {\it et al.},
\Journal{\PRL}{76}{4675}{1996}.
 
\bibitem{instanton}G.~'t~Hooft, \Journal{\PRL}{37}{8}{1976};
{\it ibid.}, \Journal{\PRD}{14}{3432}{1976}.

\bibitem{future}D.~J.~Muller and S.~Nandi (in preparation).

\end{thebibliography}
\end{document}